# Towards Key Performance Indicators of Research Infrastructures

Jana Kolar[1], Marjan Cugmas[2], Anuška Ferligoj[2]

[1]CERIC-ERIC, S.S. 14 - km 163,5 in AREA Science Park - 34149 - Basovizza, Trieste – ITALY, jana.kolar@ceric-eric.eu, corresponding author

[2]University of Ljubljana, Faculty of Social Sciences, Kardeljeva ploščad 5, 1000 Ljubljana - SLOVENIA

## Abstract

In 2018, the European Strategic Forum for research infrastructures (ESFRI) was tasked by the Competitiveness Council, a configuration of the Council of the EU, to develop a common approach for monitoring of Research Infrastructures' performance. To this end, ESFRI established a working group, which has proposed 21 Key Performance Indicators (KPIs) to monitor the progress of the Research Infrastructures (RIs) addressed towards their objectives. The RIs were then asked to assess their relevance for their institution. The paper aims to identify the relevance of certain indicators for particular groups of RIs by using cluster and discriminant analysis. This could contribute to development of a monitoring system, tailored to particular RIs.

To obtain a typology of the RIs, we first performed cluster analysis of the RIs according to their properties, which revealed clusters of RIs with similar characteristics, based on to the domain of operation, such as food, environment or engineering. Then, discriminant analysis was used to study how the relevance of the KPIs differs among the obtained clusters. This analysis revealed that the percentage of RIs correctly classified into five clusters, using the KPIs, is 80%. Such a high percentage indicates that there are significant differences in the relevance of certain indicators, depending on the ESFRI domain of the RI. The indicators therefore need to be adapted to the type of infrastructure. It is therefore proposed that the Strategic Working Groups of ESFRI addressing specific domains should be involved in the tailored development of the monitoring of pan-European RIs.

## 1 Introduction

Performance monitoring is an important element of the management of publicly funded institutions and policy interventions. It is often implemented using a system of key performance indicators (KPIs), which describe how well an institution or a program is achieving its objectives. When designed and implemented properly, they allow monitoring of progress, enable evidence-based decision-making, and aid in the development of future strategies. They can also significantly contribute to the successful communication of results and achievements, and thus to the financial sustainability of institutions and programs, as well as to increased transparency. In addition, they play a role in the evaluation of socio-economic return.[1]

Despite their importance, a recent questionnaire of European relevance revealed that although all of the 36 responding Research Infrastructures (RIs) agreed that they should have KPIs, only half of them actually have them.[2] In addition, only 3 RIs reported that their KPIs adhered to the RACER criteria, i.e., that they are relevant, accepted, credible, easy to monitor and robust.[3] This result points to one of the key issues related to KPIs – considerable effort and specific knowledge is required to develop a set of high quality KPIs that an RI can use to track its progress towards objectives. Since all RIs need KPIs, the joint development of a set of KPIs to be adopted voluntarily by RIs that share some of the objectives, would be sensible. In fact, the ERIC FORUM project, an EC co-funded project that brings together more than half of the pan-European research infrastructures in operation, has such a development among its tasks.[4]

In addition to the performance management of RIs themselves, policy makers have identified the need for performance monitoring of their support to research infrastructures. The stakeholder consultation on the long-term sustainability of RIs that the European Commission undertook in 2017, revealed that there is a need to develop appropriate KPIs, which would serve as a good basis for achieving sustainable monitoring and governance of Research and Data Infrastructures. Following this input, the European Commission has included in its action plan the need to assess the quality and impact of the RI and its services, by developing a set of key performance indicators, based on excellence principles.[5] This was supported by the Competitiveness Council, which in its conclusions invited Member States and the Commission to develop within the framework of ESFRI a common approach for monitoring RI's performance, and invited the Pan-European Research Infrastructures, on a voluntary basis, to include it in their governance and explore options to support this through the use of key performance indicators".[6]

In response to the invitation, the European Strategic Forum for Research Infrastructures (ESFRI) established a Working Group on Monitoring (WG) in 2018, which is tasked, among other things, with developing a core set of KPIs that could be applied across different RIs.

In order to propose KPIs that can be applied across RIs, the WG reviewed the objectives that are most commonly shared among the RIs. A review of the objectives of more than half of the pan-European RIs in operation revealed that, while several of them share objectives, such as delivery of education and training, enhancing collaboration in Europe and outreach to the public, only one objective, achieving scientific excellence, is shared by all of them. This already raises concerns as to whether a core system of KPIs, applicable across RIs, can be proposed. This is not a surprising finding, due to the diversity of the RIs. Some of them offer access to their facilities, based on merit, while others offer open access to their resources, such as data, tissues or museum collections. Some of them are single sited, while others are distributed over a number of member states. Their scientific fields also vary, from physics and engineering to social sciences and humanities.

---

[1] https://www.ceric-eric.eu/2018/08/30/key-performance-indicators-of-research-infrastructures/
[2] https://www.ceric-eric.eu/2018/11/05/key-performance-indicators-of-research-infrastructures-2/
[3] https://ec.europa.eu/info/sites/info/files/file_import/better-regulation-toolbox-41_en_0.pdf
[4] https://www.eric-forum.eu
[5] https://ec.europa.eu/info/sites/info/files/research_and_innovation/research_by_area/documents/swd-infrastructures_323-2017.pdf
[6] Council conclusions on "Accelerating knowledge circulation in the EU", 29 May 2018, http://data.consilium.europa.eu/doc/document/ST-9507-2018-INIT/en/pdf

The WG nevertheless identified the main objectives shared among RIs, developed a set of KPIs and asked the RIs to assess their relevance for their institution by assigning values 1 to 4 (1 – not relevant, 4 – highly relevant). Thirty-nine pan-European RIs and a further 10 RIs of European Relevance responded to the invitation.

Considering the amount of data available, we decided to use a statistical approach to elucidate whether certain types of infrastructure could be identified based on the replies, which would enable further targeted refinement of the proposed indicators. Furthermore, the relevance of various KPIs for each group of RIs could be assessed, which is important for the development of the tailored monitoring system.

## 2  Data

Two sets of variables were used in the analysis: the characteristics of the RIs and the ratings RIs gave for the relevance of the key performance indicators, from 1 to 4, 4 being highly relevant. The proposed indicators used are presented in Table 1, numbered from 1-20. Indicators number 7, Number of MSC and PhD thesis, and number 21, Revenues, were excluded from the study since it was clear from the comments of the respondents that they interpreted them in different ways. The responding RIs were assigned descriptors describing their properties (Annex 1). ESFRI areas, as determined in the ESFRI Roadmap 2018,[7] are as follows:

1 - Energy

2- Environment

3- Health and Food

4 – Physical Sciences and Engineering

5 - Social and Cultural Innovation

6 - e-RI

The other characteristics were whether the RI is:

- Pan European (1) or a national facility (0)
- In operation (1) or not (0)
- Resource RI, such as data, collections (1) or not (0)

---

[7] http://roadmap2018.esfri.eu

- Facility RI, enabling access to physical infrastructure (1) or not (0)
- Distributed over many locations (1) or single-sited (0)
- e-Infrastructure (1) or not (0)

The descriptors were based on the ESFRI 2018 roadmap descriptions of the projects, self-declarations of RIs and other web-page descriptions of the RIs.

*Table 1: Proposed KPIs by the ESFRI Working Group on Monitoring included in the Questionnaire to the RIs*

| Number | Indicator name |
|---|---|
| 1 | Experimental time available or size of resources database |
| 2 | Number of proposals/user requests, or Number of registered users of data, services (for resources RIs) |
| 3 | Number of granted proposals/ accepted users, or Number of logins/month; number of downloads, number of studies or services (for resources RIs) |
| 4 | Number of publications |
| 5 | Proportion of publications in top 10% in comparable field |
| 6 | Person-hours for staff receiving training |
| 8 | No of hours/no of participants in training events or through on-line services |
| 9 | Number of members (from other EU countries) |
| 10 | Share of users and publications per EU country |
| 11 | Share of publications/co-publications with industry |
| 12 | Share of revenues from economic activities in the annual accounts |
| 13 | Number of events organised for target groups and number of participants |
| 14 | Number of times the RI is mentioned in media articles, radio or TV broadcasts or web-based media |
| 15 | Website popularity and level of social media engagement |
| 16 | Number of publicly available data sets used externally |
| 17 | Participation by RIs in policy related events, committees & advisory boards |
| 18 | Number of times the RI or its projects are cited in policy related publications |
| 19 | Share of research projects with one or more partners outside the EU |
| 20 | Training in an international context (participant-days) |

## 3  Analysis

The RIs are described by binary variables. The percentages of 'yes' (1) answers are therefore given in Table 2. The areas covered by the RIs vary considerably, with the largest number of responding RIs belonging to the Physical Sciences and Engineering (PSE, 34.7), while only two of the RIs are e-Infrastructures. Furthermore, a large majority of the responding RIs are pan-European (79.6%), in operation (75.5%), distributed (69.4%).

The distributions and some statistics of the ordinal variables describing the performance indicators are given in Table 3. It can be concluded that there are very different mean values of the considered indicators. Not surprisingly, the most relevant ones address the size of resources (1) or the objective of scientific excellence (2-4). In fact, a recent review of the Statutes of RIs organized as an ERIC (European Research Infrastructure Consortia) revealed that contributing to scientific excellence was the only objective, which is shared by all the ERICs.[8] The outcome indicator from this group, Proportion of publications in top 10% in comparable field (no. 5) scored a bit lower, since it is a more controversial indicator. It is calculated based on articles published in peer-reviewed journals included in the databases Web of Science or Scopus. It is therefore less relevant for RIs in the fields of social sciences and humanities, which publish extensively in other types of literature. Among the three indicators with the lowest relevance scores, two address economic activities or cooperation with industry, namely Share of publications/co-publications with industry (no. 11) and Share of revenues from economic activities in the annual accounts (no. 12). This is also to be expected, since several RIs do not have commercial activities among their objectives, progress towards which the KPIs should measure. In fact, only 55% of ERICs share the objective of pursuing technological development, innovation and knowledge transfer, and not all are necessarily linked to the commercial activities and collaboration with industry. The third indicator with a low relevance is Person-hours for staff receiving training (no. 6). The objective of training is shared by 45% of ERICs, however, while training of users is a relevant indicator (no. 8), Person-hours for staff receiving training was generally perceived as a less relevant one and it is proposed to be excluded from the list of proposed indicators.

*Table 2: Characteristics of the RIs*

| Variable | Frequency ("yes" answers) | Percentage ("yes" answers) |
|---|---|---|
| ESFRI (Energy) | 3 | 6.1 |
| ESFRI (Environment) | 10 | 20.4 |
| ESFRI (Health and food) | 10 | 20.4 |
| ESFRI (PSE) | 17 | 34.7 |
| ESFRI (Social and Cultural Innovations) | 7 | 14.3 |
| ESFRI (e-RI) | 2 | 4.1 |
| Pan European RI | 39 | 79.6 |
| RI in operation | 37 | 75.5 |
| Resource RI | 25 | 51.0 |
| Facility RI | 35 | 71.4 |
| Distributed RI | 34 | 69.4 |
| E--RI | 2 | 4.1 |

The research question was whether certain types of infrastructures can be identified based on the relevance of the key performance indicators. The results would answer which indicators (KPIs) are particularly relevant for a particular group of RIs, and enable further targeted refinement of the monitoring system.

---

[8] CERIC's review of the objectives of ERICs, update, https://www.ceric-eric.eu/wp-content/uploads/2019/09/Objectives-of-ERIC_Sept2019-1.pdf

*Table 3: Distributions of the key performance indicators (1 – not relevant; 4 – highly relevant)*

|  | 1 – not relevanjt | 1.5 | 2 | 2.5 | 3 | 3.5 | 4 – highly relevant | median | mean | standard deviation |
|---|---|---|---|---|---|---|---|---|---|---|
| Indicator 1 | 5 | 0 | 2 | 0 | 9 | 1 | 32 | 4 | 3.4 | 1.0 |
| Indicator 2 | 0 | 0 | 5 | 0 | 7 | 0 | 37 | 4 | 3.7 | 0.7 |
| Indicator 3 | 0 | 0 | 3 | 0 | 7 | 3 | 36 | 4 | 3.7 | 0.6 |
| Indicator 4 | 1 | 0 | 2 | 0 | 12 | 1 | 33 | 4 | 3.6 | 0.7 |
| Indicator 5 | 6 | 0 | 8 | 0 | 15 | 1 | 19 | 3 | 3.0 | 1.0 |
| Indicator 6 | 13 | 0 | 15 | 0 | 13 | 0 | 8 | 2 | 2.3 | 1.0 |
| Indicator 8 | 7 | 0 | 11 | 0 | 14 | 0 | 17 | 3 | 2.8 | 1.1 |
| Indicator 9 | 3 | 0 | 6 | 0 | 12 | 1 | 27 | 4 | 3.3 | 0.9 |
| Indicator 10 | 8 | 0 | 11 | 0 | 19 | 1 | 10 | 3 | 2.7 | 1.0 |
| Indicator 11 | 8 | 0 | 21 | 0 | 12 | 0 | 8 | 2 | 2.4 | 1.0 |
| Indicator 12 | 18 | 0 | 16 | 0 | 7 | 0 | 8 | 2 | 2.1 | 1.1 |
| Indicator 13 | 0 | 0 | 3 | 1 | 18 | 0 | 27 | 4 | 3.5 | 0.6 |
| Indicator 14 | 3 | 0 | 18 | 1 | 20 | 0 | 7 | 3 | 2.6 | 0.8 |
| Indicator 15 | 1 | 0 | 11 | 2 | 19 | 0 | 16 | 3 | 3.0 | 0.8 |
| Indicator 16 | 7 | 0 | 8 | 1 | 13 | 0 | 20 | 3 | 2.9 | 1.1 |
| Indicator 17 | 3 | 0 | 8 | 1 | 17 | 0 | 20 | 3 | 3.1 | 0.9 |
| Indicator 18 | 5 | 1 | 10 | 0 | 21 | 0 | 12 | 3 | 2.8 | 0.9 |
| Indicator 19 | 2 | 0 | 9 | 3 | 12 | 0 | 23 | 3 | 3.2 | 0.9 |
| Indicator 20 | 9 | 1 | 13 | 0 | 13 | 0 | 13 | 3 | 2.6 | 1.1 |

To obtain the typology of the infrastructure organizations (RIs), we first performed cluster analysis of the infrastructure organizations according to their binary characteristics (Annex 1). This procedure reveals clusters of organizations with similar characteristics. To obtain these clusters of organizations, we first calculated Euclidean distances among them and then, based on these distances, we applied the Ward agglomerative hierarchical clustering procedure.[9] The obtained hierarchical clustering is presented by the dendrogram in Figure 1.

Secondly, to study how the relevance of the key performance indicators (Table 1) differs among the obtained clusters, discriminant analysis was used. Here, a discriminant variable, which is the sum of weighted key performance indicators, is estimated in such a way that it best discriminates the clusters. By this criterion the weights or loadings are obtained. The indicators with larger loadings best discriminate the clusters of organizations. Therefore, the groups in the discriminant analysis are the clusters obtained by cluster analysis and the variables that discriminate them are the key performance indicators.

---

[9] Ward, J. H., Jr. (1963), "Hierarchical Grouping to Optimize an Objective Function", *Journal of the American Statistical Association*, 58, 236–244.

## 3.1 Cluster analysis

The obtained dendrogram (Figure 1) shows two possible partitions of RIs: into two clusters and into five clusters, according to their ESFRI domains and other RIs' characteristics.

In the case of the 2-cluster solution, the two clusters contain the following infrastructures:
- CLUSTER 1: RIs from the ESFRI domains Environment, Health and Food, Social and Cultural Innovation and e-Infrastructures.
- CLUSTER 2: RIs from the ESFRI domains Physical Sciences and Engineering and Energy.

The 5-cluster solution results in the following clusters of RIs:
- CLUSTER 1: RIs from the ESFRI domain Environment
- CLUSTER 2: RIs from the ESFRI domain Health and Food
- CLUSTER 3: National RIs from the ESFRI domain Physical Sciences and Engineering
- CLUSTER 4: Pan-European RIs from the ESFRI domains Physical Sciences and Engineering and Energy.
- CLUSTER 5: RIs from the ESFRI domains Social and Cultural Innovation and e-Infrastructures.

*Figure 1: Dendrogram of infrastructure organizations*

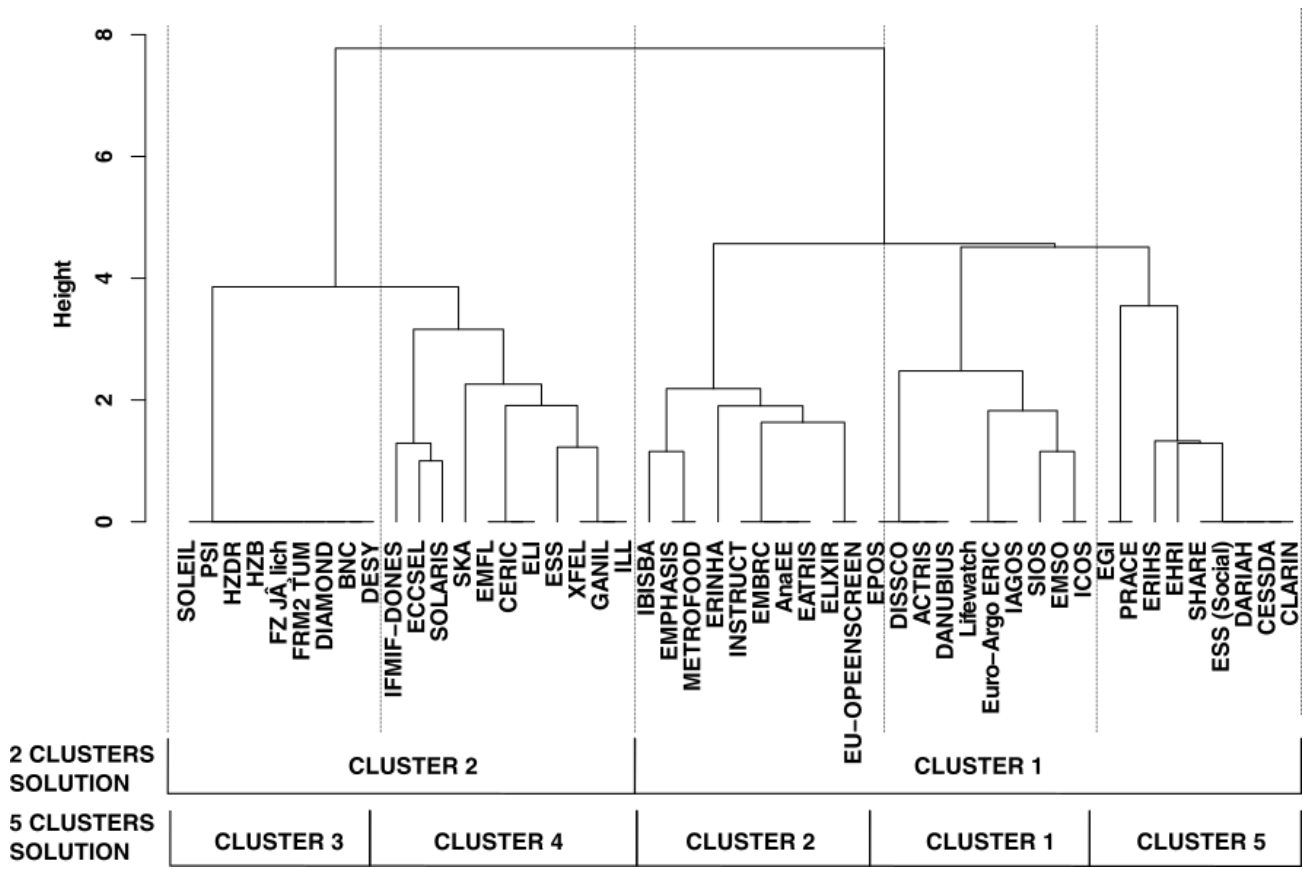

## 3.2 Discriminant analysis

Two discriminant analyses were performed. In the first, the clustering solution into two clusters is considered and 19 key performance indicators. Table 4 presents the structural loadings (these are the Pearson correlations between the obtained discriminant variable and each performance indicator). The indicators that best discriminate the two groups are highlighted in this table.

The obtained discriminant variable statistically significantly distinguishes the two obtained clusters ($F(19,29) = 2.49, p < 0.05$). Two clusters were obtained by the clustering procedure. Each organization was again classified into two clusters according to the obtained discriminant variable. Table 5 shows how well the discriminant variable that is defined by the key performance indicators, predicts the two clusters. The percentage of correctly classified organizations, based on the KPIs, is 87.76 %.

*Table 4: Structural loadings of the obtained discriminant variable; those in bold are the indicators that best distinguish the two clusters*

| indicators | | structural loadings |
|---|---|---|
| Indicator 1 | Experimental time available or size of resources database | 0.28 |
| Indicator 2 | Number of proposals/user requests, or Number of registered users of data, services (for resources RIs) | 0.08 |
| Indicator 3 | Number of granted proposals/ accepted users, or, Number of logins/month; number of downloads, number of studies or services (for resources RIs) | 0.13 |
| **Indicator 4** | **Number of publications** | **0.36** |
| **Indicator 5** | **Proportion of publications in top 10% in comparable field** | **0.50** |
| Indicator 6 | Person-hours for staff receiving training | -0.03 |
| **Indicator 8** | **No of hours/no of participants in training events or through on-line services** | **-0.44** |
| **Indicator 9** | **Number of members (from other EU countries)** | **-0.57** |
| Indicator 10 | Share of users and publications per EU country | 0.04 |
| Indicator 11 | Share of publications/co-publications with industry | 0.16 |
| **Indicator 12** | **Share of revenues from economic activities in the annual accounts** | **0.34** |
| **Indicator 13** | **Number of events organized for target groups and number of participants** | **-0.52** |
| Indicator 14 | Number of times the RI is mentioned in media articles, radio or TV broadcasts or web-based media | -0.02 |
| **Indicator 15** | **Website popularity and level of social media engagement** | **-0.61** |
| **Indicator 16** | **Number of publicly available data sets used externally** | **-0.57** |
| **Indicator 17** | **Participation by RIs in policy related events, committees & advisory boards** | **-0.40** |
| Indicator 18 | Number of times the RI or its projects are cited in policy related publications | -0.26 |
| **Indicator 19** | **Share of research projects with one or more partners outside the EU** | **-0.32** |
| Indicator 20 | Training in an international context (participant-days) | -0.12 |

*Table 5: Classification table of the RIs into two predicted clusters*

|  |  | Predicted clusters | | |
|---|---|---|---|---|
|  |  | 1 | 2 | Sum |
| Clusters obtained by clustering procedure | 1 - RIs from the ESFRI domains Environment, Health, Food, Social and Cultural Innovation and e-Infrastructures. | 26 | 3 | 29 |
|  | 2 - RIs from the ESFRI domains Physical Sciences, Engineering and Energy. | 3 | 17 | 20 |
|  | Sum | 29 | 20 | 49 |

Figure 2 graphically presents the centroids of each cluster, which are the averages of the discriminant variable for the units of each cluster.

Highly relevant indicators for organizations for Cluster 2 (RIs from the ESFRI domains Physical Sciences and Engineering and Energy) are related to Proportion of publications in top 10% in comparable field, Number of publications and Share of revenues from economic activities in the annual accounts. These indicators are particularly important for RIs that enable access to the facilities, rather than to the data or the collections. The latter (Cluster 1: RIs from the ESFRI domains Environment, Health and Food, Social and Cultural Innovation and e-Infrastructures) have often difficulties reporting on publications based on their data, since they do not know who their users are. Economic activities are also less relevant, since they often offer unrestricted access to their resources. Less relevant are the indicators Website popularity and level of social media engagement, Number of members (from other EU countries), Number of publicly available data sets used externally, Number of events organized for target groups and number of participants, No of hours/no of participants in training events or through on-line services, Participation by RIs in policy related events, committees & advisory boards and Share of research projects with one or more partners outside the EU.

*Figure 2: Two centroids of the clusters obtained by clustering procedure on the discriminant variable*

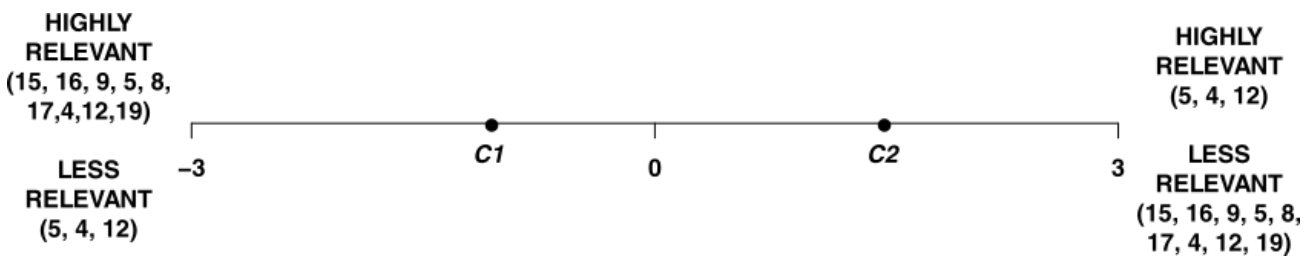

The second discriminant analysis considered the five clusters solution and the same key performance indicators. Here, four discriminant variables at most can be obtained ($min(k-1, m)$), where $k$ is the number of clusters and $m$ is the number of variables) but only one statistically significantly distinguishes the obtained five clusters (see Table 6).

Each organization was again classified into these five clusters according to the obtained discriminant variables. Table 7 shows how well the discriminant variables that are defined by the key performance indicators, predict the five clusters. The percentage of correctly classified organizations is 79.59 %. Figure 3 graphically presents the centroids of each of the five clusters on the first obtained discriminant variable. The discriminant variable is described at its ends by the most important indicators that distinguish five clusters.

*Table 6: Structural loadings of the obtained discriminant variables; those in bold are the indicators that best distinguish the five clusters*

| indicators | | Structural loadings | | | |
|---|---|---|---|---|---|
| | | D1 | D2 | D3 | D4 |
| 1 | Experimental time available or size of resources database | -0.20 | 0.95 | 0.04 | -0.36 |
| **2** | **Number of proposals/user requests, or Number of registered users of data, services (for resources RIs)** | **0.56** | -0.15 | -0.27 | 0.11 |
| 3 | Number of granted proposals/ accepted users, or, Number of logins/month; number of downloads, number of studies or services (for resources RIs) | 0.10 | -0.37 | 0.43 | -0.35 |
| **4** | **Number of publications** | **0.72** | -0.45 | 0.45 | 0.15 |
| **5** | **Proportion of publications in top 10% in comparable field** | **0.44** | -0.43 | 0.55 | -0.07 |
| **6** | **Person-hours for staff receiving training** | **0.30** | -0.10 | -0.21 | -0.30 |
| **8** | **No of hours/no of participants in training events or through on-line services** | **-0.40** | 0.18 | -0.16 | 0.28 |
| **9** | **Number of members (from other EU countries)** | **-0.56** | -0.05 | 0.78 | 0.43 |
| **10** | **Share of users and publications per EU country** | **0.34** | -0.16 | 0.43 | -0.16 |
| **11** | **Share of publications/co-publications with industry** | **0.69** | 0.21 | -0.54 | 0.57 |
| 12 | Share of revenues from economic activities in the annual accounts | -0.24 | -0.04 | -0.24 | -0.39 |
| **13** | **Number of events organized for target groups and number of participants** | **0.34** | -0.66 | 0.02 | -0.15 |
| 14 | Number of times the RI is mentioned in media articles, radio or TV broadcasts or web-based media | 0.04 | 0.31 | -0.25 | -0.19 |
| **15** | **Website popularity and level of social media engagement** | **-0.36** | -0.47 | -0.04 | -0.55 |
| **16** | **Number of publicly available data sets used externally** | **-0.93** | -0.56 | -0.20 | -0.33 |
| 17 | Participation by RIs in policy related events, committees & advisory boards | 0.08 | 0.48 | 0.09 | 0.72 |
| 18 | Number of times the RI or its projects are cited in policy related publications | 0.05 | -0.45 | -0.43 | 0.34 |
| **19** | **Share of research projects with one or more partners outside the EU** | **-0.72** | 0.01 | -0.14 | 0.10 |
| 20 | Training in an international context (participant-days) | 0.03 | 0.03 | -0.05 | -0.14 |

*Table 7: Classification table of the RIs into five predicted clusters*

|  |  | Predicted clusters | | | | | |
|---|---|---|---|---|---|---|---|
|  |  | 1 | 2 | 3 | 4 | 5 | Sum |
| Clusters obtained by clustering procedure | 1 - Environment | 9 | 1 | 0 | 0 | 0 | 10 |
|  | 2 – Health and Food | 2 | 7 | 1 | 0 | 0 | 10 |
|  | 3 – Physical sciences and Engineering (national) | 0 | 1 | 7 | 1 | 0 | 9 |
|  | 4 - Physical sciences, Engineering and Energy (pan-European) | 0 | 1 | 1 | 9 | 0 | 11 |
|  | 5 – Social and Cultural innovation | 1 | 1 | 0 | 0 | 7 | 9 |
|  | Sum | 12 | 11 | 9 | 10 | 7 | 49 |

The first discriminant variable distinguishes the best between Cluster 4 (pan-European RIs in Physical sciences and engineering and Energy) and Cluster 5 (Social and cultural innovation and e-Infrastructures). Highly relevant indicators for organizations for Cluster 4 are related to scientific excellence (Proportion of publications in top 10% in comparable field, Number of publications) and economic activities (Share of revenues from economic activities in the annual accounts and Share of publications/co-publications with industry). The less relevant indicators for this cluster are those indicators related to outreach to the public (Website popularity and level of social media engagement and Number of events organized for target groups and number of participants), the re-use of data (Number of publicly available data sets used externally), and Number of members (from other EU countries) and No of hours/no of participants in training events or through on-line services.
The organizations in Cluster 5 evaluate these indicators in the opposite way.

*Figure 3: Five centroids of the clusters on the first discriminant variable (G1: RIs from the ESFRI domain Environment; G2: RIs from the ESFRI domain Health and Food; G3: National RIs from the ESFRI domain Physical Sciences and Engineering; G4: Pan-European RIs from the ESFRI domains Physical Sciences and Engineering and Energy, G5: RIs from the ESFRI domains Social and Cultural Innovation and e-Infrastructures)*

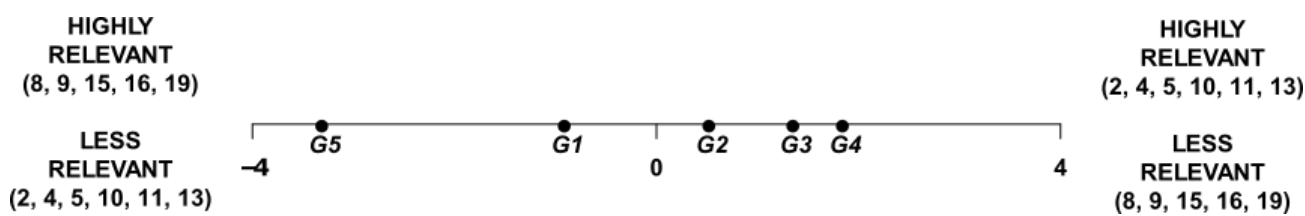

## 4 Conclusions

The research question was whether certain types of infrastructure can be identified based on the relevance of the key performance indicators. We were additionally interested in which KPIs are of particular relevance for a certain group of RIs. The results would answer whether the same indicators should be applied across all RIs and, if not, enable further targeted refinement of the proposed indicators.

We have first looked at the distributions and some statistics of the variables describing the performance indicators in order to assess whether the proposed indicators are of general relevance to the RIs and which are the most and the least relevant ones. Not surprisingly, among the five that were assessed as the most relevant, four address the size of resources or the objective of scientific excellence. This can be understandable, since a recent review of the Statutes of RIs organized as an ERIC (European Research Infrastructure Consortia) revealed that contributing to scientific excellence was the only objective, which is shared by all the ERIC. The outcome indicator from this group, Proportion of publications in top 10% in comparable field (no. 5) scored a bit lower, since it is a more controversial indicator, with lower relevance for RIs in the fields of social sciences and humanities, which publish extensively in other types of literature other than peer-reviewed papers. Among the three indicators with the lowest relevance scores, two address economic activities or cooperation with industry, namely Share of publications/co-publications with industry (no. 11) and Share of revenues from economic activities in the annual accounts (no. 12). This is also to be expected, since several RIs do not have commercial activities among their objectives, progress towards which the KPIs should measure. In fact, only 55% of ERICs share the objective of pursuing technological development, innovation and knowledge transfer, and not all of these are necessarily linked to the commercial activities. The third indicator with a low relevance is Person-hours for staff receiving training (no. 6). The objective of training is shared by 45% of ERICs, however, while training of users is a relevant indicator (no. 8), Person-hours for staff receiving training was generally perceived as a less relevant one and it is proposed to be excluded from the list of indicators.

To obtain the typology of the infrastructure organizations, we first performed cluster analysis of the infrastructure organizations according to their properties, such as their ESFRI domain, whether they offer access to facilities or resources, or are e-infrastructures, whether they are distributed among several locations, whether they are in operation and whether they are national or pan-European RIs. This procedure gives clusters of organizations with similar characteristics. Five clusters were obtained, based on the ESFRI domain and on whether the RIs are national or pan-European RIs.

Secondly, discriminant analysis was used to study how the relevance of the key performance indicators differs among the obtained clusters. In the discriminant analysis the groups are the clusters obtained by cluster analysis and the variables that discriminate them are the key performance indicators. The analysis revealed that the percentage of correctly classified organizations in the five clusters, using the key performance indicators, is 80%. Such a high percentage indicates that there are significant differences in the relevance of certain indicators, among the ESFRI domains of the RI.

We then tried to identify the indicators that best discriminate the clusters. Using discriminant analysis for five clusters, four discriminant variables can be obtained but only one statistically significantly distinguishes the five clusters. It best distinguishes between Cluster 4 (pan-European RIs in Physical sciences and Engineering and Energy) and Cluster 5 (Social and cultural innovation and e-Infrastructures). The organizations in Cluster 4 evaluate as highly relevant indicators related to scientific excellence, as expressed through publications and economic activities, while the more relevant indicators for Cluster 5 include some of the indicators related to outreach to the public, the reuse of data and similar. This can be explained by the fact that several of the RIs in the domain of social sciences and humanities (SSH) offer open access to their collections. They have difficulties in tracking the publications of their users. The other indicator in this category, Proportion of publications in top 10% in comparable field, is based on publications included in databases such as Web of Science and Scopus, while researchers in the SSH domain often publish their work in publications that are not included in these databases, rendering the indicator less relevant. In contrast to the domains of physical sciences, engineering and energy, cooperation with industry is also largely not among their objectives, so the KPIs addressing it are less or not relevant at all.

Based on the analysis presented here, it can be concluded that there is a significant difference in the relevance of the proposed KPIs among the groups of RIs, primarily based on their domain of operation. The indicators, therefore, need to be adapted to the type of infrastructure. It is proposed that the Strategic Working Groups of ESFRI that address specific domains, such as energy, environment or health, should be involved in the further development of the monitoring of pan-European RIs.

## Acknowledgement

The study is undertaken in the frame of ACCELERATE project, co-funded by the European Union Framework Programme for Research and Innovation Horizon 2020, under grant agreement 731112.

# Annex 1: Properties of the RIs*

| Research Infrastructure (RI) | ESFRI Area | Pan European RI | RI in operation | Resource RI | Facility RI | Distributed RI | e-RI |
|---|---|---|---|---|---|---|---|
| ECCSEL - European Carbon Dioxide Capture and Storage Laboratory Infrastructure | 1 | 1 | 1 | 0 | 1 | 1 | 0 |
| IFMIF-DONES - International Fusion Materials Irradiation facility - DEMO Oriented Neutron Source | 1 | 1 | 0 | 0 | 1 | 0 | 0 |
| SOLARIS - European SOLAR Research Infrastructure for Concentrated Solar Power | 1 | 1 | 0 | 0 | 1 | 1 | 0 |
| ACTRIS - Aerosols, Clouds and Trace gases Research Infrastructure | 2 | 1 | 0 | 1 | 1 | 1 | 0 |
| DANUBIUS - International Centre for Advanced Studies on River-Sea Systems | 2 | 1 | 0 | 1 | 1 | 1 | 0 |
| DISSCO - Distributed System of Scientific Collections | 2 | 1 | 0 | 1 | 1 | 1 | 0 |
| EMSO - European Multidisciplinary Seafloor and water-column Observatory | 2 | 1 | 1 | 1 | 1 | 1 | 0 |
| EPOS - European Plate Observing System | 2 | 1 | 0 | 1 | 1 | 1 | 0 |
| Euro-Argo - European contribution to the international Argo Programme | 2 | 1 | 1 | 1 | 0 | 1 | 0 |
| IAGOS - In-service Aircraft for a Global Observing System | 2 | 1 | 1 | 1 | 0 | 1 | 0 |
| ICOS - Integrated Carbon Observation System | 2 | 1 | 1 | 1 | 1 | 1 | 0 |
| Lifewatch - e-Infrastructure for Biodiversity and Ecosystem Research | 2 | 1 | 1 | 1 | 0 | 1 | 0 |
| SIOS - The Svalbard Integrated Arctic Earth Observing System | 2 | 1 | 1 | 0 | 1 | 1 | 0 |
| AnaEE - Infrastructure for Analysis and Experimentation on Ecosystems | 3 | 1 | 1 | 1 | 1 | 1 | 0 |
| EATRIS - European Advanced Translational Research Infrastructure in Medicine | 3 | 1 | 1 | 1 | 1 | 1 | 0 |
| ELIXIR - A distributed infrastructure for life-science information | 3 | 1 | 1 | 1 | 0 | 1 | 0 |
| EMBRC - European Marine Biological Resource Centre | 3 | 1 | 1 | 1 | 1 | 1 | 0 |
| EMPHASIS - European Infrastructure for Multi-scale Plant Phenomics and Simulation | 3 | 1 | 0 | 1 | 1 | 1 | 0 |
| ERINHA - European Research Infrastructure on Highly Pathogenic Agents | 3 | 0 | 1 | 0 | 1 | 1 | 0 |
| EU-OPEENSCREEN - European Infrastructure of Open Screening Platforms for Chemical Biology | 3 | 1 | 1 | 1 | 0 | 1 | 0 |
| IBISBA - European Industrial Biotechnology Innovation and Synthetic Biology Accelerator | 3 | 1 | 0 | 0 | 1 | 1 | 0 |
| INSTRUCT - Integrated Structural Biology Infrastructure | 3 | 1 | 1 | 1 | 1 | 1 | 0 |
| METROFOOD - Infrastructure for Promoting Metrology in Food and Nutrition | 3 | 1 | 0 | 1 | 1 | 1 | 0 |
| BNC - Budapest Neutron Centre | 4 | 0 | 1 | 0 | 1 | 0 | 0 |

| Name | | | | | | | |
|---|---|---|---|---|---|---|---|
| CERIC - Central European Research Infrastructure Consortium | 4 | 1 | 1 | 0 | 1 | 1 | 0 |
| DESY - German Electron Synchrotron | 4 | 0 | 1 | 0 | 1 | 0 | 0 |
| DIAMOND - Diamond Light Source | 4 | 0 | 1 | 0 | 1 | 0 | 0 |
| ELI - Extreme Light Infrastructure | 4 | 1 | 1 | 0 | 1 | 1 | 0 |
| EMFL - European Magnetic Field Laboratory | 4 | 1 | 1 | 0 | 1 | 1 | 0 |
| ESS – European Spallation Source | 4 | 1 | 0 | 0 | 1 | 0 | 0 |
| FRM2 TUM - Forschungs-Neutronenquelle Heinz Maier-Leibnitz | 4 | 0 | 1 | 0 | 1 | 0 | 0 |
| FZ Jülich – Forschungzentrum Jülich | 4 | 0 | 1 | 0 | 1 | 0 | 0 |
| GANIL (Spiral 2) - Système de Production d'Ions Radioactifs en Ligne de 2e génération | 4 | 1 | 1 | 0 | 1 | 0 | 0 |
| HZB – Helmholtz-Zentrum Berlin | 4 | 0 | 1 | 0 | 1 | 0 | 0 |
| HZDR – Helmholz-Zentrum Dresden Rossendorf | 4 | 0 | 1 | 0 | 1 | 0 | 0 |
| ILL - Institut Max von Laue - Paul Langevin | 4 | 1 | 1 | 0 | 1 | 0 | 0 |
| PSI - The Paul Scherrer Institute | 4 | 0 | 1 | 0 | 1 | 0 | 0 |
| SKA - Square Kilometre Array | 4 | 1 | 0 | 1 | 0 | 0 | 0 |
| SOLEIL – Synchrotron SOLEIL | 4 | 0 | 1 | 0 | 1 | 0 | 0 |
| XFEL - European X-Ray Free-Electron Laser Facility | 4 | 1 | 1 | 0 | 1 | 0 | 0 |
| CESSDA - Consortium of European Social Science Data Archives | 5 | 1 | 1 | 1 | 0 | 1 | 0 |
| CLARIN - Common Language Resources and Technology Infrastructure | 5 | 1 | 1 | 1 | 0 | 1 | 0 |
| DARIAH - Digital Research Infrastructure for the Arts and Humanities | 5 | 1 | 1 | 1 | 0 | 1 | 0 |
| EHRI - European Holocaust Research Infrastructure | 5 | 1 | 0 | 1 | 0 | 1 | 0 |
| E-RIHS - European Research Infrastructure for Heritage Science | 5 | 1 | 1 | 1 | 1 | 1 | 0 |
| ESS - European Social Survey | 5 | 1 | 1 | 1 | 0 | 1 | 0 |
| SHARE - Survey of Health, Ageing and Retirement in Europe | 5 | 1 | 1 | 1 | 0 | 1 | 0 |
| EGI - The EGI Foundation | 6 | 1 | 1 | 0 | 0 | 1 | 1 |
| PRACE - Partnership for Advanced Computing in Europe | 6 | 1 | 1 | 0 | 0 | 1 | 1 |